# A Hybrid Forecast of Exchange Rate based on Discrete Grey-Markov and Grey Neural Network Model

**Gol Kim [a], Ri Suk Yun [b]**

([a] Center of Natural Science, University of Sciences, Pyongyang, DPR Korea,
E-mail: golkim124@yahoo.com
[b] *Foreign Economic General Bureau, Pyongyang, DPR Korea*)

**Abstract:** We propose a hybrid forecast model based on discrete grey-fuzzy Markov and grey –neural network model and show that our hybrid model can improve much more the performance of forecast than traditional grey-Markov model and neural network models. Our simulation results are shown that our hybrid forecast method with the combinational weight based on optimal grey relation degree method is better than the hybrid model with combinational weight based minimization of error-squared criterion.

**Keywords:** Financial forecasting, Hybrid forecast, Discrete Grey-Fuzzy Markov, Grey –Neural Network.

## 1. Introduction

Investors have been trying to find a way to predict exchange rates and stock price accurately, but haven't been obtained the good results.Recently, artificial intelligence techniques like artificial neural networks (ANN) and genetic algorithms (GA), and wavelet transform have been applied to this area. However, the above-mentioned concern method showed that ANN had some limitations in learning the patterns because foreign exchange rates data has tremendous noise and complex dimensionality. Moreover, the sheer quantity of exchange rates data sometimes interferes with the learning of patterns.
Hybrid forecast is a well-established and well-tested approach for improving the forecasting accuracy. Therefore, the importance of hybrid forecast methods has steadily increased and it acts still on time series forecasting. Hybrid forecast system using rough sets and artificial neural networks has been proposed by M.Duraira, K. Meena (2011) and hybrid forecast method by ARIMA and neural network model Has been researched by Reza Askari Moghadam, Mehrnaz Keshmirpour (2011). Aobing Sun, Yubo Tan, Dexian Zhang (2008) have researched the hybrid forecast model based on BP neural network.
Hybrid Forecast Model of ARIMA and Neural Network has been given by Šterba Ján, Hil'ovsk Katarína (2010) and stacked heterogeneous neural networks for time series forecasting has been suggested by Florin Leon, Mihai Horia Zaharia(2010).
Tarek Abouedahab, Mahumod Fakhreldin (2011) have studied the forecast of stock market indices using hybrid genetic algorithm/ particle swarm optimization with perturbation term.
Improvement method in forecasting accuracy using the hybrid model of ARFIMA and feed forward neural network has been proposed by Cagdas Hakan Aladag, Erol Egrioglu, Cem Kadilar (2012).
Forecast of future stock close price using based on hybrid ANN Model of functional link fuzzy logic neural model has been researched by Kumaran Kumar. J, Kailas (2012).
Forecasting of foreign exchange rate by grey- chaotic forecast hybrid has been researched by Kim Gol (2008) and forecasting of stock price and index of oil by grey- fractal **forecast** hybrid model has been studied by Kim Gol (2009).
Zhu Jiang-liang (2006) has studied the forecasting of the electric power load based on the grey theory and BP neural network and Su Bo (2006) has considered the comparison and research of gain production forecasting with methods of GM (1, N) grey system and BPNN. Method of the wavelet neural network in the forecast of stock market has been suggested by,Yao Hong-Xing, Shong Zhao-han, Chen Hong-xiang (2002) and Genetic algorithms approach to feature discretization in artificial neural networks for the forecast of stock price has been researched by Kyong-jac Kim, Ingoo Han (2000).

This paper deals with the development of an improved forecast model using grey-Markov chain models and the grey-neural network and its application to financial time series forecasting. Our method includes *discrete grey- fuzzy weight Markov chain model, grey-neural network model, The hybrid model with combinational weight based minimization of error-squared criterion, hybrid model with combinational weight based on optimal grey relation degree method.*

We demonstrate the advantage of our method compared with the traditional grey-Markov or neural network methods through forecasting simulations.

## 2. Discrete grey- Fuzzy weight Markov model

### 2.1. Grey forecasting model

Grey forecasting model (GM) has three basic operation such as ;

(i) accumulated generation, (ⅱ) inverse accumulated generation, and (ⅲ) grey modeling.

The grey forecasting model uses the operations of accumulated generation to build differential equations. Intrinsically speaking, it has the characteristics of requiring less data.

The GM (1, 1) grey model, i.e., a 1- variable first-order grey model, is summarized as follows

Step 1: Selection of the original sequence;

$$x^{(0)} = (x^{(0)}(1), x^{(0)}(2), \cdots x^{(0)}(i), \cdots, x^{(0)}(N)) \tag{5}$$

where $x^{(0)}(i)$ is the time series data at time $i$.

Step 2: Accumulated generating;

Based on the time sequence $x^{(0)}$, a new sequence $x^{(1)}$ is given by the accumulated generating operation (AGO), where
$x^{(1)} = (x^{(1)}(1), x^{(1)}(2), \cdots x^{(1)}(i), \cdots, x^{(1)}(N))$ and is $x^{(1)}(k)$ derived as follows:

$$x^{(1)}(k) = \sum_{i=1}^{k} x^{(0)}(i) \tag{6}$$

Step 3: Making the differential equation;

The first-order differential equation holds true:

$$\frac{dx^{(1)}}{dt} + ax^{(1)} = u \tag{7}$$

Step 4: Obtaining the forecast model; We have by (7)

$$\hat{x}^{(1)}(k+1) = (x^{(0)}(1) - \frac{u}{a})e^{-ak} + \frac{u}{a} \tag{8}$$

$$\hat{x}^{(0)}(k+1) = \hat{x}^{(1)}(k+1) - \hat{x}^{(1)}(k) \tag{9}$$

where

$$[a, u]^T = (B^T B)^{-1} B^T y_N \tag{10}$$

$$B = \begin{bmatrix} -0.5(x^{(1)}(1) + x^{(1)}(2)) & 1 \\ -0.5(x^{(1)}(2) + x^{(1)}(3)) & 1 \\ \vdots & \vdots \\ -0.5(x^{(1)}(n-1) + x^{(1)}(N)) & 1 \end{bmatrix} \tag{11}$$

$$y_N = (x^{(0)}(2), x^{(0)}(3), \cdots x^{(0)}(i), \cdots, x^{(0)}(N))^T \tag{12}$$

$\hat{x}^{(1)}(k+1)$ is the predicted value of $x^{(1)}(k+1)$ at time $k+1$,

The authors used the posterior test to evaluate the accuracy of the grey forecasting. The forecasting error wear defined as $q^{(0)}(k) = x^{(0)}(k) - \hat{x}^{(0)}(k), k = 1, 2, \cdots, N$, the mean and the standard deviation of the forecasting error is $\bar{q}$ and $S_2$. The mean and the standard deviation of initial time series are

$$\bar{m} = \frac{\sum_{k=1}^{N} x^{(0)}(k)}{N} \quad , \quad S_1 = \sqrt{\sum_{k=1}^{N}(x^{(0)}(k)-\bar{m})^2 / N-1}$$

The posterior ratio $C$ is derived by dividing $S_2$ by $S_1$. The lower $C$ is, the better the model is. The posterior ratio can indicate the change rate the forecasting error. Probability of small is defined as $P = prob.\{|q^{(0)}(k) - q| \leq 0.6745 S_1\}$, where $k = 2, 3, \cdots, N$. $P$ This shows the probability that the relative bias of the forecasting error is lower than 0.6745. $P$ is commonly required to be larger than 0.95. The pairs of the forecasting indicators $P$ and $C$ can characterize four grades of forecasting accuracy, as shown Table 1.

It is called that above-mentioned grey forecast model is traditional grey model.

Table 1. The Grades of Forecasting Accuracy

| Grade | Forecasting index | |
|---|---|---|
| | $P$ | $C$ |
| Good | >0.95 | <0.35 |
| Qualified | >0.8 | <0.5 |
| Just | >0.7 | <0.65 |
| Unqualified | ≤0.7 | ≥0.65 |

## 2.2. Non homogenous discrete grey model

Let $x^{(0)} = (x^{(0)}(1), x^{(0)}(2), \cdots x^{(0)}(i), \cdots, x^{(0)}(N))$ be an original sequence and $x^{(1)} = (x^{(1)}(1), x^{(1)}(2), \cdots x^{(1)}(i), \cdots, x^{(1)}(N))$ be accumulated generating sequence by AGO. Where $x^{(1)}(k)$ is given such as:

$$x^{(1)}(k) = \sum_{i=1}^{k} x^{(0)}(i)$$

Then, we define non-homogenous discrete grey model (DGM) such as [2];

$$\begin{cases} \hat{x}^{(1)}(k+1) = \beta_1 \hat{x}^{(1)}(k) + \beta_2 \hat{x}^{(1)}(0) + \beta_3 k + \beta_4 \\ \hat{x}^{(1)}(1) = \xi \end{cases}$$

where $\hat{x}^{(1)}(k)$ fitting value of original is sequence and $\beta_1, \beta_2, \beta_3, \beta_4$ are parameters of system. By least square method, the parameters of system $\beta = [\beta_1, \beta_2, \beta_3, \beta_4]$ are given by

$$\beta = [\beta_1, \beta_2, \beta_3, \beta_4] = (B^T B)^{-1} B^T Y$$

Here, $B = \begin{bmatrix} x^{(1)}(1) & x^{(0)}(1) & 1 & 1 \\ x^{(1)}(2) & x^{(0)}(2) & 2 & 1 \\ \vdots & \vdots & \vdots & \vdots \\ x^{(1)}(n-1) & x^{(0)}(n-1) & n-1 & 1 \end{bmatrix} \quad Y = \begin{bmatrix} x^{(1)}(2) \\ x^{(1)}(3) \\ \vdots \\ x^{(1)}(n) \end{bmatrix}$

The algorithm of non-homogenous discrete grey model (DGM) is given such as;

Step 1; Find the parameters of system $\beta = [\beta_1, \beta_2, \beta_3, \beta_4]$.

Step 2; Put $\hat{x}^{(1)}(1) = \xi$ and find the simulation values $\hat{x}^{(0)}(k), k = 2, 3, \cdots, n$.
$\hat{x}^{(0)}(k)$ is the function of $\xi$.

Step 3; Calculate $Q = \sum_{k=1}^{n} [\hat{x}^{(0)}(k) - x^{(0)}(k)]^2$. Q is the function of $\xi$.

Step 4; Put $\frac{dQ}{d\xi} = 0$ and calculate the value $\xi$ which Q have to take minimum.

Step 5; Calculate the value $\hat{x}^{(1)}(k)$ corresponding to $\xi$ obtained from step 4.

Step 6; calculate the forecast value $\hat{x}^{(0)}(k), k = 2,3,\cdots,n$ by inverse accumulated generating operator (IAGO). That is, $\hat{x}^{(0)}(k) = \hat{x}^{(1)}(k+1) - \hat{x}^{(1)}(k)$.

**2.3. Fuzzy weight Markov model**

We assume that $\hat{X}_t$ is the fitting curve which is obtained through forecasting for the time series $Y_t$ by non homogenous discrete grey model.

By considering the actual meaning of the original sequence, we have generated a new time series $Z_t = (Y_t - \hat{X}_t)/Y_{t-1}$ $(t = 2,3,\cdots,N)$ which the fitting curve $\hat{X}_t$ takes as reference

Here $Y_{t-1}$ is one step of time delay value.

In accordance with the distribution of the random sequence $Z_t$, it is divided by $k$ pieces of state and is taken its partition value. That is,

$$e_1 = [m_0, m_1], e_2 = [m_1, m_2],,\cdots, e_{k-1} = [m_{k-2}, m_{k-1}], e_k = [m_{k-1}, m_k]$$

Here $m_i \geq m_{i-1}$.

A one-step transition probability $P$ is associated with each possible transition from stat $e_i$ to state $e_j$, and $P$ can be estimated using $P_{ij} = M_{ij}/M_i$ $(i, j = 1,2,\cdots,m)$. $M_i$ means the numbers of divided pieces whose residuals are stat $e_i$, and $M_{ij}$ is number of transition from $e_i$ to $e_j$ that have occurred by passed one step. These $P_{ij}$ values can be presented as a transition matrix $R$.

Then, we accept the statistical quantity such as;

$$P_{0j} = \sum_{i=1}^{k} M_{ij} / \sum_{i=1}^{k} M_i, \quad \chi^2 = 2\sum_{i=1}^{k}\sum_{j=1}^{k} M_{ij} \left|\log\frac{P_{ij}}{P_{0j}}\right|$$

Then, when $N$ is comparatively large, the statistical quantity $\chi^2$ is according to $\chi^2$-distribution with the degree of freedom $(k-1)^2$.

When giving the confidence degree $\alpha$, if $\chi^2 > \chi^2_\alpha(m-1)^2$, then we confirm that the random sequence $Z_t$ have Markov's property, unless, the sequence haven't Markov's property.

Suppose $U$ is the range which random variable of Markov chain takes the value.

We construct the fuzzy state set $S_1, S_2, \cdots, S_l$. If for arbitrary $u \in U$ the condition

$$\sum_{m=1}^{l} \mu_{S_m}(u) = 1$$

is satisfied, and then $\mu_{S_m}(u)$ is called the membership degree of the fuzzy state $S_M$ for numerical value $U$.

Suppose that $\mu_{S_i}(Z_t) \cdot \mu_{S_j}(Z_{t+1})$ is the fuzzy state transition coefficient from the state $S_i$ to $S_j$ when time is turned from $t$ to $t+1$.

Then,

$$a_{ij} = \sum_{t=1}^{N-1} \mu_{S_i}(Z_t) \cdot \mu_{S_j}(Z_{t+1})$$

is called fuzzy transition frequency number from the state $S_i$ to the state $S_j$.

When the state $Z_t$ belongs to $S_i$ with degree of member $\mu_{S_i}(Z_t)$ and belongs to $S_j$ with membership degree $\mu_{S_j}(Z_t)$, the transition order is only expressed by the product between membership degrees $\mu_{S_i}(Z_t) \cdot \mu_{S_j}(Z_{t+1})$.

Owing to the fuzzy transition probability from the state $S_i$ to the state $S_j$ is denoted by

$$P_{ij} = \frac{a_{ij}}{\sum_{j=1}^{\infty} a_{ij}}, \ ((i,j=1,2,\cdots,M))$$

, the time series time series which we are going to establish is given such as;

$$\hat{Y}_t = \hat{X}_t + \sum_{i=1}^{k} \mu_{S_i}(Z_{t-1}) \sum_{j=1}^{k} \frac{1}{2}(m_{i-1} + m_i) P_{ij} Y_{t-1}, \ t=1,2,\cdots,N$$

Here, $\hat{X}_t$ is the predicted value obtained by using the discrete grey model.
The forecasting model obtained by above methods is called the discrete grey -fuzzy weight Markov model. We denote this model by DGM-FMarkov.

## 3. The forecast of JPY/USD exchange rate based on the discrete grey -fuzzy weight Markov model.

We take JPY/USD exchange rate of total 278 barter period from September 1994 to July 1997. The data of this paper is taken from G. Shinarist, P. Waimiso (1997). The concrete data is omitted. We assume that $\hat{y}_k$ is the forecasted value of the original data obtained by DGM model in $k$ time. The curve $\hat{y}_k = \hat{x}_{k+1}^{(0)}$ deflects the total change tendency of the original data.
Based on the real phase of the sample data and considering the real meaning, let divide the state 6' th item, that is,

$$\otimes_i = [\otimes_{1i}, \otimes_{2i}] = [\widehat{y_k} + a_i x_k^{(0)}, \widehat{y_k} + b_i x_k^{(0)}] \ (i=1,2,\cdots,6)$$

where $a_i = (a_1, a_2, \cdots, a_m) = (-0.09, -0.025, -0.01, 0, 0.01, 0.025)$,

$b_i = (b_1, b_2, \cdots, b_m) = (-0.025, -0.01, 0, 0.01, 0.025, 0.09)$

$\otimes_1 : [\widehat{y_k} - 0.09 x_k^{(0)}, \widehat{y_k} - 0.025 x_k^{(0)}]$ , $\otimes_2 : [\widehat{y_k} - 0.025 x_k^{(0)}, \widehat{y_k} - 0.01 x_k^{(0)}]$

$\otimes_3 : [\widehat{y_k} - 0.01 x_k^{(0)}, \widehat{y_k}]$ , $\otimes_4 : [\widehat{y_k}, \widehat{y_k} + 0.01 x_k^{(0)}]$

$\otimes_5 : [\widehat{y_k} + 0.01 x_k^{(0)}, \widehat{y_k} + 0.025 x_k^{(0)}]$ , $\otimes_6 : [\widehat{y_k} + 0.025 x_k^{(0)}, \widehat{y_k} + 0.09 x_k^{(0)}]$

where $\hat{y}_k$ is the predict value of the foreign exchange rate obtained by DGM model in $k$ time. The state shows the degree of deviation of the initial data deviated from the curve $\hat{y}_k$. Where $\otimes_1$ denote that the disadvantage is putted within 2~9% range and $\otimes_2$ disadvantage is putted within 1~2.5% range. And $\otimes_3$ and $\otimes_4$ denote that the profit and disadvantage are putted within 1% range and $\otimes_5$ denote that the profit is putted within 1~2.5% range and $\otimes_6$ denote that the profit is putted within 2.5% range.
The states are $\otimes_1, \otimes_2, \otimes_3, \otimes_4, \otimes_5, \otimes_6$. The numbers of sample point lay at each state are respectively $n_1 = 73, \ n_2 = 42, n_3 = 29, n_4 = 19, n_5 = 44, n_6 = 71$. The 1-step state matrix composed by each state is given by

$$(n_{ij})_{6\times 6} = \begin{bmatrix} 53 & 14 & 3 & 0 & 2 & 1 \\ 13 & 18 & 9 & 2 & 0 & 0 \\ 4 & 9 & 12 & 3 & 1 & 0 \\ 0 & 2 & 6 & 8 & 3 & 0 \\ 1 & 0 & 0 & 9 & 13 & 10 \\ 0 & 0 & 0 & 0 & 23 & 48 \end{bmatrix}$$

The last state transition of the initial sample is uncertain. Therefore, when calculating the state transition matrix, last data have to exclude. From the formula (15), we can obtain

$$M_{0j} = [0.2626, \quad 0.1511, \quad 0.1043, \quad 0.0683, \quad 0.1583, \quad 0.2554]^T$$

From the formula (16), we can obtain the state transition matrix by

$$(M_{ij})_{6\times 6} = \begin{bmatrix} 0.7260 & 0.1918 & 0.9411 & 0 & 0.0274 & 0.0137 \\ 0.3095 & 0.4286 & 0.2143 & 0.0476 & 0 & 0 \\ 0.1319 & 0.1303 & 0.4138 & 0.1034 & 0.0345 & 0 \\ 0 & 0.1053 & 0.3158 & 0.4211 & 0.1579 & 0 \\ 0.0227 & 0 & 0 & 0.2645 & 0.2955 & 0.5000 \\ 0 & 0 & 0 & 0 & 0.3239 & 0.7324 \end{bmatrix}$$

Calculating the statistical quantity, we have

$$\chi^2 = 2\sum_{i=1}^{m}\sum_{j=1}^{m} n_{ij} \left| \log \frac{M_{ij}}{M_{0j}} \right| \approx 280.89$$

Taking the degree of confidence $\alpha = 0.01$ and inspecting the tables, then we have $\chi^2_{0.01}(25) = 44.3$, $\chi^2 \geq \chi^2_{0.01}(25)$. Therefore, the 1-step transition matrix $(M_{ij})_{6\times 6}$ then we confirm that the sequence is coincidence to Markov's property, unless, the sequence is not coincidence to Markov's property. Then, based on this transition matrix, we can predict the foreign exchange rates which would be putted on the next 1-period or $k$-period. Therefore, the forecast value of the exchange rates of the 279'th period is given by

$$\hat{x}^{(0)}_{279} = [\hat{y}_{278} + \frac{1}{2}(0.01\hat{x}^{(0)}_{278} + 0.025\hat{x}^{(0)}_{278})p_{65} + \frac{1}{2}(0.025\hat{x}^{(0)}_{278} + 0.09\hat{x}^{(0)}_{278})p_{66}] = 117.49$$

From the step-by-step formula the forecast value of the stock price of the period which would be putted on afterward the state 279'th period is given by

$$\hat{x}^{(0)}_{k+1} = \hat{y}_k + \frac{1}{2}\hat{x}^{(0)}_k [0.115 p_{66}(k+1) + 0.025 p_{65}(k+1)$$
$$- 0.01 p_{63}(k+1) - 0.035 p_{62}(k+1) - 0.115 p_{61}(k+1)], k \geq 279$$

Accuracy was the most important criterion, followed by the cost savings generated from improved decisions. In particular, execution issues such as ease of interpretation and ease of use were also highly rated. In this study, there are three criteria used to evaluate forecasting models.

The first evaluate criteria is mean square error, MSE:

$$MSE = \sum_{k=1}^{N} (\hat{x}^{(0)}(k) - x^{(0)}(k))^2 / N \tag{17}$$

where $\hat{x}^{(0)}(k)$ is the predicted value at time $k$, $x^{(0)}(k)$ is the actual value at time $k$, and $N$ is the number of forecasts.

The second criterion is mean absolute error, MAE:

$$MAE = \sum_{k=1}^{N} |e^{(0)}(k)| / N \equiv \sum_{k=1}^{N} |\hat{x}^{(0)}(k) - x^{(0)}(k)| / N \tag{18}$$

The third criterion is mean absolute percent error, MAPE:

$$MAPE = \frac{1}{N}\sum_{i=1}^{N} |q^{(0)}(k)| \times 100(\%) \equiv \frac{1}{N}\sum_{i=1}^{N} \left| \frac{\hat{x}^{(0)}(k) - x^{(0)}(k)}{x^{(0)}(k)} \right| \times 100(\%) \tag{19}$$

The fourth criterion is Theil Coefficient, $\mu$:

$$\mu^2 = \frac{[\sum (x^{(0)}(k) - \hat{x}^{(0)}(k))^2]/N}{[\sum (\hat{x}^{(0)}(k))^2]/N} \tag{20}$$

If $x^{(0)}(k) = \hat{x}^{(0)}(k)(1 \leq k \leq N)$, i.e., forecast value is coinciding with actual value perfectly, we have $\mu^2 = 0$. Therefore, if $\mu^2$ is approached to $0$, then forecast method become finer.

## 4. Grey Neural Network

The Grey GM (1, 1) model was established forecast model by using new data accumulated generator. The accumulated generator data can make weak the randomness a certain degree and can easily find the data change rule. The grey GM (1, 1) model has the advantage which demands the small sample data.

The neural network has the capacity of self-learning, nonlinear mapping, and parallel distribution processing, etc.

These tow forecast method have already used in foreign exchange rate predicting.

The system of foreign exchange rate forecast is a complicated system which have the great randomness and there are many index influenced in that system.

Combining grey system idea and neural network, we can construct the grey neural network (GNN) and sufficiently demonstrate the advantage of the model by using the modeling method of grey system with the small data and the characters of nonlinear mapping of neural network.

Thus, we can more enhance forecast accuracy.

Here we discuss the method of combing the GM model and neural network model. There are three kinds of forecasting model structure. That is, Parallel grey neural network (PGNN), series grey neural network (SGNN), and inlaid grey neural network (IGNN).

### 4.1. Parallel grey neural network (PGNN)

PGNN uses grey model and neural network to predict separately, then uses neural network to combine the predicting results. Fig. 1 shows the principle scheme of PGNN

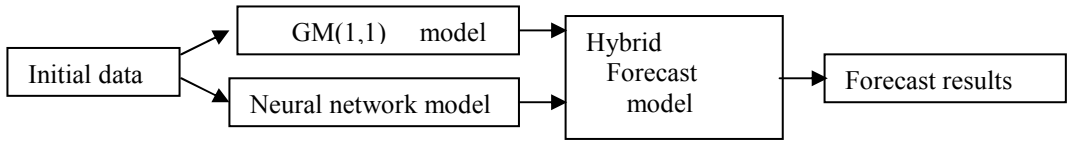

Fig.1 Parallel grey neural network model

PGNN model is essential is the hybrid model. The aim of making the PGNN model is to decrees the randomness of data and avoids data lack used by single model, and enhances the forecast accuracy degree by using data offered from various kinds methods synthetically.

On the ground of hybrid forecast principle, there are various hybrid methods that is, arithmetic mean hybrid method, geometric mean hybrid method, harmonic mean hybrid method, etc. Those hybrid formulas are given by

$$\hat{x}(k) = w_1 \hat{x}_1(k) + w_2 \hat{x}_2(k), k = 1, 2, \cdots, N$$
$$\hat{x}(k) = \hat{x}_1(k)^{w_1} + \hat{x}_2(k)^{w_2}, k = 1, 2, \cdots, N$$
$$\hat{x}(k) = \frac{1}{w_1/\hat{x}_1(k) + w_2/\hat{x}_2(k)}, k = 1, 2, \cdots, N$$

Where $N$ is total forecast data number, $\hat{x}_1(k)$ and $\hat{x}_2(k)$ are the forecast value by GM(1,1) model and neural network model, respectively, and $w_1$ and $w_2$ are weights of two kind of forecast method. $\hat{x}(k)$ is actual value.

Now, we consider the conception of effective degree. The conception of effective degree has some rationality because it reflects the efficient of the methods.
Its idea is as follow;
We put

$$A(k) = 1 - \left|\frac{x(k) - \hat{x}(k)}{x(k)}\right| = 1 - \left|\frac{x(k) - (w_1\hat{x}_1(k) + w_2\hat{x}_2(k))}{x(k)}\right|$$

Then, $A(k)$ is the accuracy series of hybrid forecast. We denote the mean value $E$ and mean square deviation of $A(k)$ respectively by

$$E = \frac{1}{N}\sum_{k=1}^{N} A(k), \; \sigma = \frac{1}{N}(\sum_{k=1}^{N}(A(k)-E)^2)^{1/2}.$$

We define effective degree of hybrid forecast method by
$S = E(1-\sigma)$
If $S$ is more great, then forecast accuracy is more raised and forecast error is more safety. It says that the model is more effective.

To find the weights $w_1, w_2$, we can use optimization method by

$$\inf_{\{w_1,w_2\}} S(w_1,w_2) \equiv S = E(1-\sigma)$$

But, this method is very complicated.
Now, we would use a simple method by using of physical meaning.
Let $A_1(k)$ and $A_2(k)$ are the series of forecast accuracy by grey GM (1, 1) model method and neural network model method respectively, that is,

$$A_1(k) = 1 - \left|\frac{x(k) - \hat{x}_i(k)}{x(k)}\right|, i=1,2; k=1,2,\cdots,N$$

Let $S_i$ are the effective degree by GM (1, 1) Model and neural network model respectively. Then, we can defined weights $w_1$ and $w_2$ by

$$w_i = \frac{S_i}{\sum_{j=1}^{2} S_j} \; (i=1,2).$$

**4.2. Series grey neural network (SGNN)**

Series grey neural network employ grey model to predict, then use neural network to combine the predicting results.
Established forecast model GM(1,1) with each other data for already given the same series, then the gained forecast results are different each other.
To obtain forecast results approached to actual value, for various grey forecast results we can combine its results by using neural network. This is immediately SGNN model.
Fig. 2 shows the principle scheme of SGNN.
The input nerve element number of neural network is the number of each other GM model and the number of output nerve element number of neural network is single element. The hidden nerve element number is confirmed by test methods. The learning of neural network is progressed by error back-propagation method.
PGNN and SGNN are all essential hybrid forecast models.
Theoretically, we can prove that the forecast results of hybrid model are advantage than single GM model or neural network model.

In SGNN model, we find the combined weights of several kinds grey model by using the nonlinear matching ability of neural network. The combination of SGNN is nonlinear, the other side, the combination of PGNN is linear.

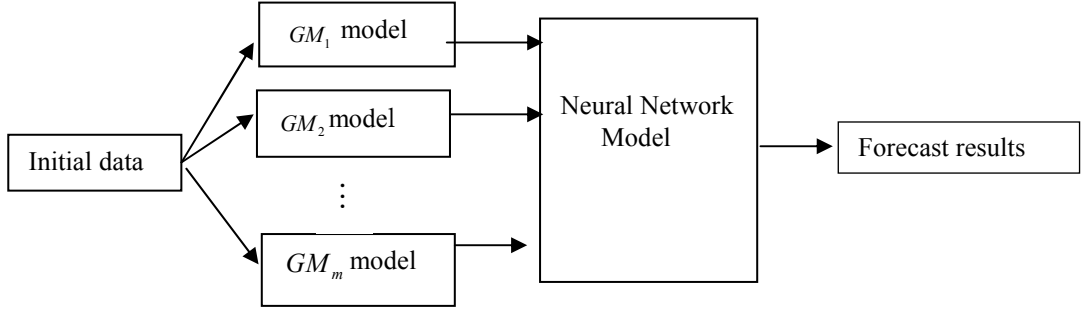

Fig.2 Series grey neural network model

### 4.3. Inlaid grey neural network (IGNN)

IGNN model is built by adding a grey layer before neural input layer and a white layer after neural output layer. Fig. 3 shows the principle scheme of IGNN

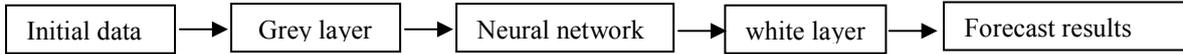

Fig.3 Inlaid grey neural network model

The action of grey layer is to weaken the randomness of initial data. Generally, the role of grey layer is to make training sample of neural network with new data gained from 1-AGO for initial data. In that case, the approximation in nonlinear excitation function of neural network is easy and studying time of network is very short, and convergence process is quick and forecast accuracy is raised because accumulated operator data have monotone increasing tendency.

## 5. The hybrid forecast method

### 5.1 The determination of combinational weight based minimization of error-squared criterion

In the actual problem, we can predict for forecasting problem by using the various forecast methods. Generally speaking, the forecast result forecasted by the distinctive forecast method is different each other. To obtain the more improved forecast result, it is necessary to apply the various kinds of hybrid forecast methods.

Let $x^{(0)}(k)$ $(k=1,2,\cdots,N)$ are an actual observation value and $\hat{x}_i(k)$ $(k=1,2,\cdots,p)$ are the predict value by the $i'$ th method. Let $w_i$ is the weight of the $i'$ th method and $e_i(k) = x^{(0)}(k) - \hat{x}_i(k)$ Is the forecast error. From this notation, we have

$$x^{(0)}(k) = \sum_{i=1}^{p}(w_i \hat{x}_i(k) + e_i(k)) \ (k=1,2,\cdots,N).$$

Now, let $\hat{w}_i$ is the calculated value of the weight and $\hat{x}^{(0)}(k)$ is the hybrid forecast value. Then we can obtain

$$\hat{x}^{(0)}(k) = \sum_{i=1}^{p} \hat{w}_i \hat{x}_i(k)$$

To find the optimal weight $w_i^*$, we can use various optimization methods. For example, find optimal weight $w_i^*$ for optimization problem;

$$\begin{cases} J[w_i^*] = \min_{w_i} J[w_i] = \sum_{k=1}^{N}(x^{(0)}(k) - \sum_{i=1}^{p} w_i \hat{x}_i(k))^2 \\ \sum_{i=1}^{p} w_i = 1, \ w_i \geq 0, \ i = 1, 2, \cdots, p \end{cases} \quad (21)$$

To obtain the solution of optimization problem (21), we can apply various optimization methods, for example, quadratic programming, genetic algorithm, neural network method, etc.

For instance, $\hat{x}_1^{(0)}(k)$ and $\hat{x}_2^{(0)}(k)$ are the predict value by two methods for the same phenomena. Then,

$$\hat{x}^{(0)}(k) = w_1 \hat{x}_1^{(0)}(k) + w_2 \hat{x}_2^{(0)}(k) = \rho \hat{x}_1^{(0)}(k) + (1-\rho) \hat{x}_2^{(0)}(k)$$

is called the hybrid forecast model for two models $\hat{x}_1^{(0)}(k)$ and $\hat{x}_2^{(0)}(k)$. Where $0 \leq \rho \leq 1$.

Now, we put $e_1 = e_1^{(0)}(k)$, $e_2 = e_2^{(0)}(k)$, from the hybrid model, we have $e = \rho e_1 + (1-\rho) e_2$.

The square deviation of $e$ is given by

$$D(e) = \rho^2 D(e_1) + (1-\rho^2) D(e_2) + 2\rho(1-\rho) \text{cov}(e_1, e_2).$$

Where $\text{cov}(e_1, e_2)$ is covariance of $e_1$ and $e_2$. To obtain least value of $D(e)$, we put $\dfrac{dD(e)}{d\rho} = 0$.

Then, we can obtain the least value of $D(e)$ by

$$\rho^* = \frac{D(e_1) - \text{cov}(e_1, e_2)}{D(e) = D(e_1) + D(e_2) - 2\text{cov}(e_1, e_2)}.$$

That is, when $\rho = \rho^*$, $D(e)$ is achieved to the least value. In the actual calculation, we can approvable that $e_1$ and $e_2$ are the mutual independence random quantity. Therefore, we have $\text{cov}(e_1, e_2) = 0$. Accordingly, we have

$$\rho^* = \frac{D(e_1)}{D(e) = D(e_1) + D(e_2)} \quad (22)$$

The final forecast model is given by

$$\hat{x}^{(0)}(k) = \rho^* \hat{x}_1^{(0)}(k) + (1-\rho^*) \hat{x}_2^{(0)}(k) \quad (23)$$

### 5.2. The determination of combinational weight based on optimal grey relation degree method

Let $\{y(k) | k = 1, 2, \cdots, N\}$ be original value of time series, $\hat{y}_j(k)$ ($i = 1, 2, \cdots, m$, $k = 1, 2, \cdots, N$) be predicted value of the $j$'th forecast method in $k$ time ($j = 1, 2, \cdots, m$, $t = 1, 2, \cdots, N$).

We put

$$\gamma_{0j} = \frac{1}{N} \sum_{t=1}^{N} \frac{\min_{1 \leq j \leq m} \min_{1 \leq t \leq N} |e_j(t)| + \rho \max_{1 \leq j \leq m} \max_{1 \leq t \leq N} |e_j(t)|}{|e_j(t)| + \rho \max_{1 \leq j \leq m} \max_{1 \leq t \leq N} |e_j(t)|}$$

Then, $\gamma_{0j}$ is called the grey relation degree of the predicted value sequence

Here $\rho \in (0,1)$ is called the identification coefficient. Generally we take $\rho = 0.5$.

$e_j(t) = y(t) - \hat{y}_j(t)$ is the predicted error of $k$ time for $i$'th forecast method.

Let $\hat{y}(t) = \sum_{j=1}^{m} w_j \hat{y}_j(t)$ be the predicted value of $y(t)$ by the hybrid forecast method.

Here $w_1, w_2, \cdots, w_m$ is the weight coefficient of $m$ kind of forecast method, which satisfies $\sum_{j=1}^{m} w_j = 1, w_j \geq 0 \ (j = 1, 2, \cdots, m)$.

The combinational weight based on the optimal grey relation degree is determined such as;

$$\max \gamma(W) = \frac{1}{N} \sum_{t=1}^{N} \frac{\min_{1\leq j\leq m} \min_{1\leq t\leq N} |e_j(t)| + \rho \max_{1\leq j\leq m} \max_{1\leq t\leq N} |e_j(t)|}{\left|\sum_{j=1}^{m} w_j e_j(t)\right| + \rho \max_{1\leq j\leq m} \max_{1\leq t\leq N} |e_j(t)|}$$

$$s.t. \begin{cases} \sum_{j=1}^{m} w_j = 1 \\ w_j \geq 0, \quad j = 1, 2, \cdots, m \end{cases}$$

By solving this optimal problem, we can determine the combinational weight $W = \{w_1, w_2, \cdots, w_m\}^T$ based on the optimal grey relation degree.

## 6. The hybrid forecast of exchange rates by DGM-FMarkov model and Grey - Neural Network

Now, let $\hat{x}_1^{(0)}(k)$ is the predict value obtained by DGM-FMarkov model and $\hat{x}_2^{(0)}(k)$ is the forecasted value obtained by inlaid grey neural network model.
We have to obtain the calculating results by DGM-FMarkov model and grey neural network model for exchange rates.
The structure of neural network is taken $4\times 4\times 1$ BP network. The transfer function of input layer and hidden layer are sigmoid type and transfer function of output is linear function.
In GNN forecast, we have applied IGNN mode. We have progressed 1-AGO smoothing handling by adding a grey layer before input layer of BP neural network and have progressed 1-IAGO reduction handling for white layer after neural output layer.
We have calculated according to formula (22), (23) by unbiased grey-Markov- grey neural network hybrid forecast value for foreign exchange rates.
The mean absolute percent error MAPE of DGM-FMarkov model is given by

$$MAPE = \frac{1}{N} \sum_{i=1}^{N} \left| \frac{\hat{x}^{(0)}(k) - x^{(0)}(k)}{x^{(0)}(k)} \right| \times 100\% \approx 3.65(\%)$$

Theil coefficient of DGM-FMarkov model is given by $\mu = 0.0053$

The mean absolute percent error MAPE by hybrid model of DGM-FMarkov model - grey neural network on the basis of the determination of combinational weight based minimization of error-squared criterion is given by

$$MAPE = \frac{1}{N} \sum_{i=1}^{N} \left| \frac{\hat{x}^{(0)}(k) - x^{(0)}(k)}{x^{(0)}(k)} \right| \times 100\% \approx 2.17(\%)$$

Then, Theil coefficient is given by $\mu = 0.0032$

The mean absolute percent error MAPE by hybrid model of DGM-FMarkov model - grey neural network on the basis of the determination of combinational weight based on optimal grey relation degree method is given by

$$MAPE = \frac{1}{N} \sum_{i=1}^{N} \left| \frac{\hat{x}^{(0)}(k) - x^{(0)}(k)}{x^{(0)}(k)} \right| \times 100\% \approx 1.85(\%)$$

Then, Theil coefficient is given by $\mu = 0.0021$

# 7. Conclusion

As known from the hybrid forecast model, its model can use the information of the original data sufficiently and absorb the characters and the advantages of two models, and avoid the limitation of single model.
The forecast accuracy of hybrid model by DGM-FMarkov model - grey neural network is more higher than the DGM-FMarkov model. Especially, The forecast accuracy of hybrid model by DGM-FMarkov - grey neural network model on the basis of the determination of combinational weight based on optimal grey relation degree method is more higher than the DGM-FMarkov-grey neural network model on the basis of the determination of combinational weight based minimization of error-squared criterion
It is shown that this model has high forecast accuracy for forecast of exchange rates with tendency and fluctuation.
 In this paper we have researched an hybrid forecast model by using DGM-FMarkov model and grey neural network forecast model and have exhibited that this method is efficient to the exchange rates forecast
In the hybrid forecast model, we can use some forecast models, fore example; not only forecast not only DGM-FMarkov forecast, grey-regression forecast, but also grey- chaotic forecast and grey-fractal forecast ([13, 14]), ect.